\documentclass{hep99}
\usepackage{epsf}
\begin{document}

\title{\boldmath
 CDF $B$ spectroscopy results: $B^{**}$ and $B^+_c$
 }

\author{G. Bauer \\ \small
(representing the CDF Collaboration)
}
%

\address{
              Laboratory for Nuclear Science, 
              Massachusetts Institute of Technology, \\
              Cambridge, MA 02139 USA
\\[3pt]
E-mail: {\tt bauerg@fnal.gov}}

\abstract{We report on two spectroscopy results from CDF.
          First, we observe  the orbitally 
          excited $B^{**}$ mesons in $B\!\rightarrow \ell D^{(*)}X$ 
          events. We find 
          $28 \pm 6 \pm 3$\% of light $B$ mesons
          produced are~$B^{**}$ states. A collective mass fit results
          in a $B_1$ mass of 
          $5.71 \pm 0.02$ GeV/$c^2$.
          Secondly,
          we observe $20.4^{+6.2}_{-5.5}$ decays of
          $B^+_c \!\rightarrow \! J/\psi \ell^+X$, with 
          a 
          $ 6.40 \pm\! 0.39 \pm\! 0.13 \,{\rm GeV}/c^2$ mass
          and $ 0.46 ^{+0.18}_{-0.16} \pm 0.03\,$ps lifetime. 
          The production rate is
          in reasonable accordance with expec\-tations.
}

\maketitle


\section{Introduction\label{sec:intro}}

The large $b$ cross section at the Tevatron make it an attractive
arena for studying $b$-hadrons. 
CDF has reported a variety of
spectroscopy results, including
the most precise mass determinations
of the $B^0_s$ \cite{CDFBs} and $\Lambda^0_b$ \cite{CDFLamb}.
Here we report results on the rare $B^+_c$, and the not rare, 
but hard to observe, $B^{**}$ states.

\section{\boldmath $B^{**}$ production \label{sec:B**}}

The  $B^{**}$ states are the 4 orbitally ($L = 1$) excited states
of the $B$ meson. 
In a relativistic light-quark model the states
 $B_1$,   $B^*_2$, $B^*_0$, and $B^*_1$ have masses
$5.719$,  $5.733$, $5.738$, and $5.757$ GeV/$c^2$~\cite{Ebert}.
Being above the $\pi$-threshold, they
decay via $B^{**} \!\rightarrow\! B^{(*)}\pi$.
The normally broad ({$\sim$}$100\,$MeV) hadronic decay width is 
expected to be suppressed ({$\sim$}$20\,$MeV) for $B_1$ and $B^*_2$
because only $L=2$ decays are allowed. 

Study of $B^{**}$'s is of interest for non-perturbative QCD models,
and for ``engineering'' $b$-flavor tagging methods \cite{SST,SSTPRD}.
$B^{**}$'s have been observed in $e^+e^-$ collisions \cite{LEP**}.
Here we report the first observation of  $B^{**}$'s 
in a hadron collider.

We use $110 \,{\rm pb}^{-1}$ of data collected in Run I.
We reconstruct 6 modes 
of the type $B \!\rightarrow\! D^{(*)} \ell X$ \cite{Dejan},
all of which have been previously documented \cite{SSTPRD} except
for the addition of $\ell^+ \overline{D}{^0}$,
$\overline{D}{^0} \!\rightarrow\! K^+\pi^-\pi^+\pi^-$.
Side-band subtractions are performed, and we effectively obtain
a pure sample of almost $10^4$ $B$'s.

$B^{**}$'s should be narrow peaks on
a broad~struc\-ture in the $B\pi$ mass. Even
after kinematic~correc\-tions ($\sim${$15$}\%) the lost $\nu$, 
as well as the~unidentified~$\gamma$~from 
$B^*$ decay, smears these peaks.~With background, it 
is then extremely difficult to identify  $B^{**}$'s.
These problems are 
ameliorated by using the 
quantity $Q \equiv m[\ell D^{(*)}\pi] - m[\ell D^{(*)}] -m[\pi]$
which compresses the broad 
$m[\ell D^{(*)}\pi]$ 
distribu\-tion (with $\ell D^{(*)} \approx B$)
into a relatively narrow
range at low $Q$.
\vspace*{-13.8cm}
{ \flushright \large
$\;$\\
FERMILAB-Conf-99/227-E \\
}
\vspace*{12.32cm}

We combine $B$'s with tracks ($p_T \!>\! 0.9\,$GeV/$c$), 
assumed to be $\pi$'s, from the primary vertex 
(impact parameter {$<\,$}$3\sigma$) to form $B^{**}$ candidates.
These $B$-$\pi$ combinations contain a variety of backgrounds
un\-cor\-related to the $B$: 
random $\pi$'s from the under\-ly\-ing event and
from multiple $\bar{p}p$ collisions.
These backgrounds may be removed by ``sideband subtrac\-tion''
methods.
The major remaining back\-ground is from pions 
from the hadronization of the $B$, which, unfortunately, 
{\it is} correlated with the $B$, and thus demands careful treatment.

\begin{figure}[t]
{\centering
  \epsfxsize=18pc 
\epsfbox{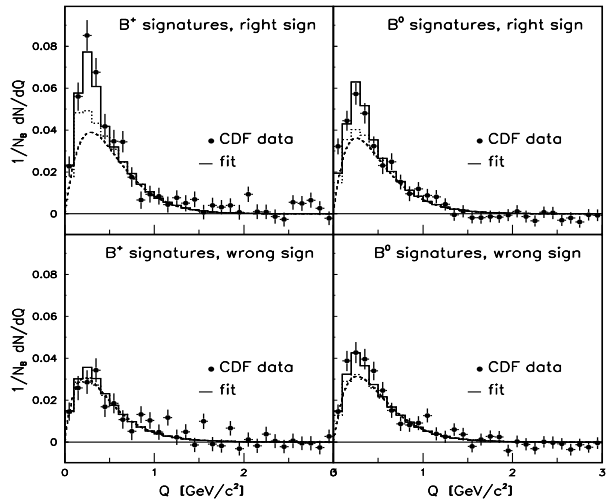}
  \caption{
     The sideband-subtracted $Q$-distributions 
divided into $B^+$/$B^0$ modes and right/wrong-sign
$B\pi^\pm$ combinations: 
data (points), fit (solid histo\-gram),
total background (dotted histogram),
and hadron\-ization back\-ground (dashed curve).
}
\label{fig:Qdist} }
\end{figure}

$B^{**}$ decays give $B^+\pi^-$ or $B^0\pi^+$ 
(``right-sign'') combinations at low-$Q$,
and not $B^+\pi^+$ or $B^0\pi^-$ (``wrong-sign'').
The $B$-$\pi$ $Q$-distributions,
divided into $B^+$ and $B^0$ mesons 
and into right/wrong-sign categories, are shown in Fig.~\ref{fig:Qdist}. 
The data (points) show
a clear right-sign excess, but $B^+$ and $B^0$ behave differently
and the wrong-sign background peaks in the same $Q$-region.
The $B^{**}$ signal is entangled with the hadronization background
which also favors the right-sign at low $Q$-values
(the basis for our ``same side tagging'' \cite{SST,SSTPRD}).
Thus, one can not expose a $B^{**}$ signal by 
subtracting the ``wrong-sign'' $Q$-distributions
from the ``right-sign'' ones.


We model the hadronization $Q$-distributions
by 2-parameter functions
inspired by {\sc PYTHIA}~\cite{pythia},
and impose the {\it relative} 
right/wrong-sign hadron\-ization asymmetry
from the simulation.
We fit the data for $B^{**}$ signal
plus this hadronization model.\footnote{Other small backgrounds,
such as $B^{**}_s$, are included.
The fit accounts for the important sample composition
issues of cross-talk between $B^+$ and $B^0$ decays and $B^0$-mixing.} 
The specific {\it shape} of the hadronization background,
as well as its overall norm\-alization,
and the amount of any $B^{**}$ signal
are free to float in the fit.

The solid histogram in Fig.~\ref{fig:Qdist} shows the
fit, with the dotted histogram showing the total background
and the dashed curve is the hadronization compo\-nent.
The excess above the total background (dot\-ted) is the $B^{**}$ signal,
which is even in the wrong-sign events.
$B^0$-mixing moves events between right-sign $B^0$'s 
and  wrong-sign $B^0$'s, creating an 
apparent asymmetry between
the $B^{**}$ signal in $B^+$'s and $B^0$'s. There is a small amount
of cross-talk between $B^+$ and $B^0$ reconstructions
({\it e.g.} if the $\pi^-$ is lost 
from $D^{*-} \!\rightarrow\! \overline{D}{^0}\pi^-$),
which shifts $B^{**}$'s diagonally in Fig.~\ref{fig:Qdist},
{\it e.g.}, right-sign  $B^+$ to wrong-sign $B^0$.

The fit results in a  $B\pi$ excess from which we find
that  $B^{**}$ states are
$28 \!\pm\! 6 \!\pm\! 3$\% of  light $B$ meson production.
The distributions of Fig.~\ref{fig:Qdist} are clearly inadequate
to distinguish the $B^{**}$ states, but we can use the mass
splitting of Ref.~\cite{Ebert} and fit 
the $Q$-distribution for the collective $B^{**}$  mass. We quote the result 
in terms of the mass of the lowest state, $B_1$, 
as $5.71 \pm 0.02 \,(stat. + syst.)$ GeV/$c^2$. \cite{Dejan} 

\section{\boldmath $B^+_c$  production \label{sec:Bc}}

The $B^+_c$ is the ground state 
of $c\bar{b}$ mesons.
It is novel as a bound state of two {\it different} heavy quarks,
and is an interesting test for bound-state models.
CDF has previously searched for the $J/\psi \pi^+$ decay,
and set upper limits \cite{CDFPsiPi}.
We extend the search to the higher rate semileptonic
mode $B^+_c \!\rightarrow\! J/\psi \ell^+\nu$ \cite{CDFBc}.

We use {$\sim$}$200,000$ $J/\psi \!\rightarrow\! \mu^+\mu^-$ events 
($p_T(\mu)$ above $\sim\!\!1.5\,$GeV/$c$) fully contained in the
Si-$\mu$vertex detector (for precision vertexing).
A 3rd track is add\-ed to the  $J/\psi$ compatible
(Prob($\chi^2$)$>1$\%) with its vertex, and within
a cone of $90^\circ$. 
The proper time for the $J/\psi+$track system must be more
than $ 60 \mu$m.
This yields 6530 (1055) candidates
satisfying electron (muon) fiducial cuts. Lepton identification 
criteria applied to the 3rd 
track reduced the sample to 23 (14) 
electron (muon) candidates.

$B^+_c$ background comes from two general
classes:  $J/\psi+$fake lepton, 
and  $J/\psi+$real, but uncorrelated, lepton.
Fake leptons arise from misidentified had\-rons 
($e^-$-misidentification, decay-in-flight,\ldots),
and uncorrelated leptons from $\gamma$-conversions
or from the semileptonic decay of a second $b$-hadron.
These backgrounds are determined from data measure\-ments
extrapolated via Monte Carlo to the $J/\psi \ell^+$ sample.
We find $8.6\pm2.0$  ($12.8\pm2.4$) 
background events in the
electron (muon) sample \cite{CDFBc}.

\begin{figure}
{\centering
\mbox{
\epsfxsize=16pc 
\epsfbox{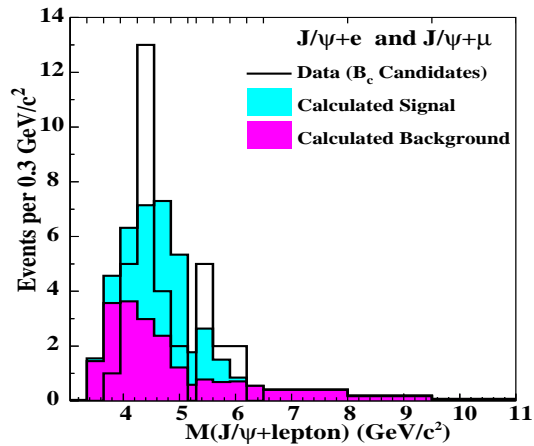}
}
  \caption{
       The $J/\psi+\ell^+$ mass distributions of data, and the
calculated background and signal.
}
\label{fig:Bcmass} }
\end{figure}

A likelihood fit of the $J/\psi+\ell^+$ mass distri\-bu\-tions 
($e^+$ and $\mu^+$ separated but 
fit simultaneously), with 
calculated backgrounds and a  $B^+_c$ component,
yields a  $B^+_c$ signal of $20.4^{+6.2}_{-5.5}$ mesons
($4.8\sigma$ sig\-nificance).
The results are shown in Fig.\ref{fig:Bcmass}.

\begin{figure}[t]
{\centering
\mbox{ 
  \epsfxsize= 8.7pc %
\epsfbox{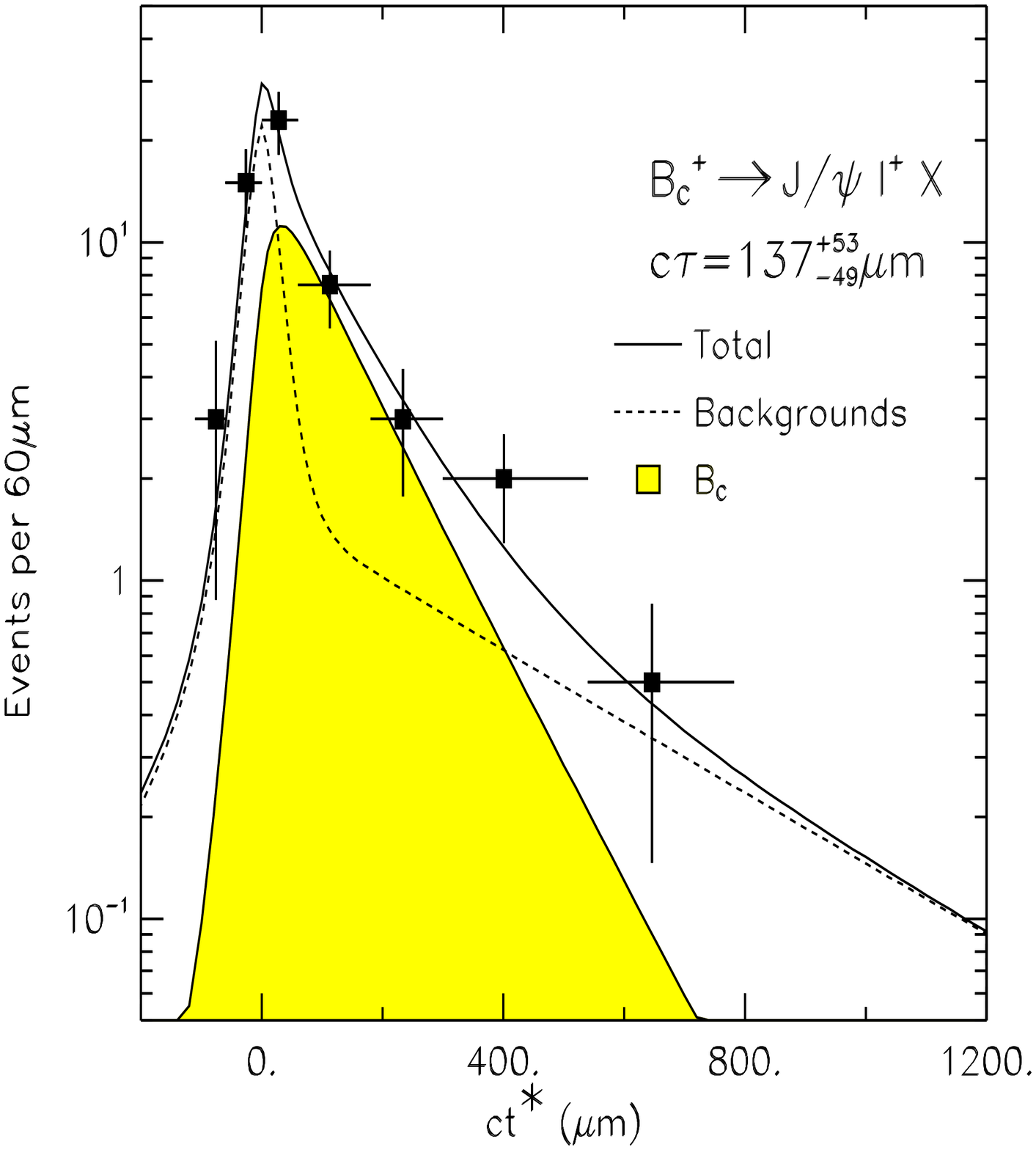}
  \epsfxsize= 9.5pc %
\epsfbox{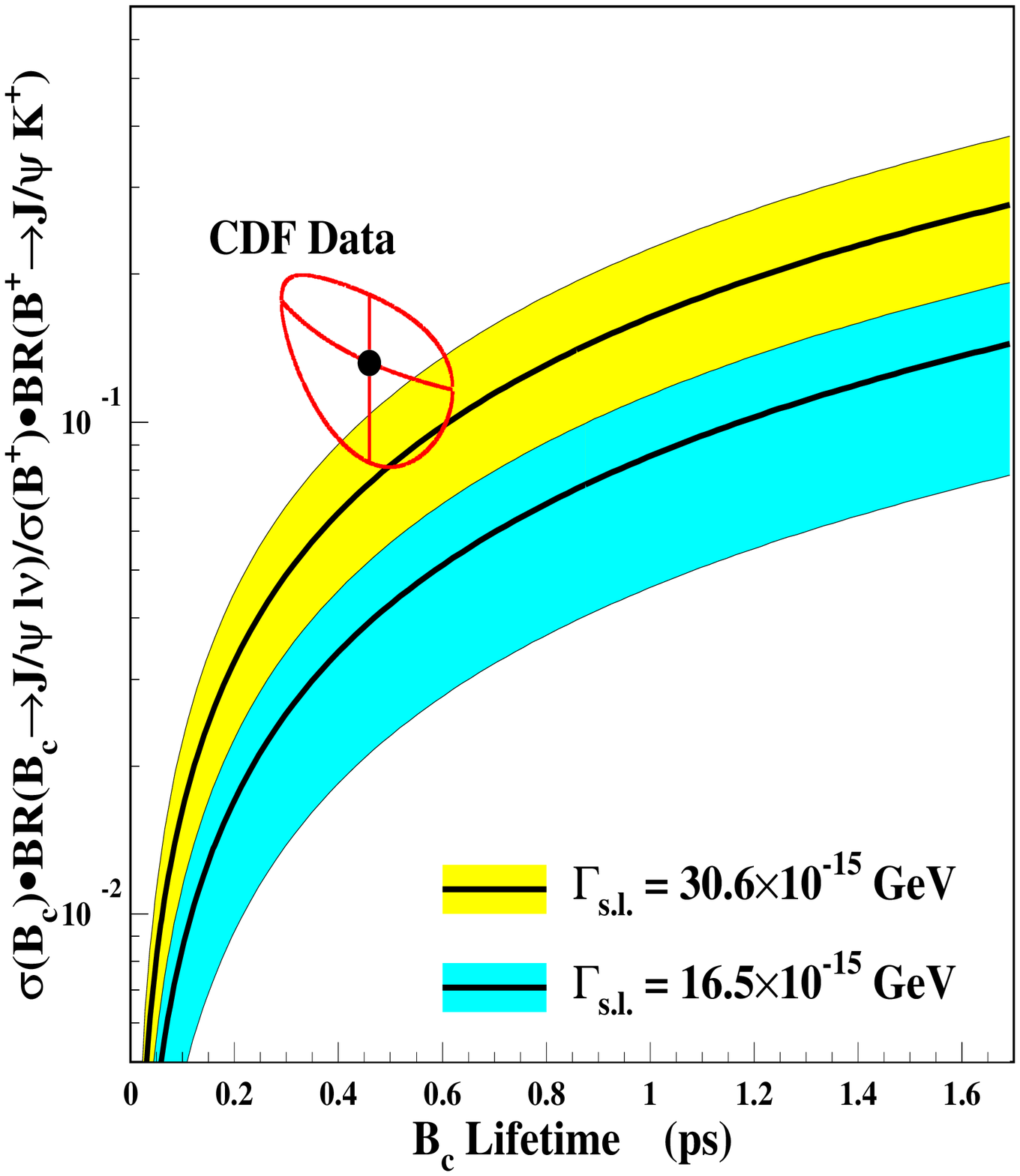}
}
  \caption{
        Left:  Proper time distribution of the data with lifetime fit.
        Right: Relative branching ratio ${\cal R}(J/\psi \ell^+\nu)$
               as a function of $B^+_c$ lifetime.
}
\label{fig:life} }
\end{figure}

This sample may be used to extract several  $B^+_c$ properties.
Although the missing $\nu$ greatly reduces sensitivity 
to the $B^+_c$ mass, we find from our fits a value of 
$ 6.40 \!\pm\! 0.39 \!\pm\! 0.13 \,{\rm GeV}/c^2$.
We release~the $60 \, \mu$m ``lifetime'' cut, correct for the
missing $\nu$ and resolution effects, and fit the lifetime distribution 
of Fig.~\ref{fig:life} to obtain
$ \tau(B^+_c) \!=\!0.46 ^{+0.18}_{-0.16} \pm 0.03\,$ps.

We can determine 
the cross-section$\times$branch\-ing-fraction,
$\sigma_{Bc} {\cal B}(B^+_c \!\rightarrow\! J/\psi \ell^+\nu )$,
from the event yield.
We do so, however, relative to the similar $B^+_u \!\rightarrow\! J/\psi K^+$
decay since many experimental sys\-tem\-atics cancel in the ratio.
We find: 

\noindent
\begin{eqnarray}
{\cal R}(J/\psi \ell^+\nu) \equiv 
  \frac{\sigma_{Bc}\times {\cal B}(B^+_c \!\rightarrow\! J/\psi \ell^+\nu ) }
       {\sigma_{Bu}\times {\cal B}(B^+_u \!\rightarrow\! J/\psi K^+)  } = 
\nonumber \\
             13.2 ^{+4.1}_{-3.7} \,(stat) \pm 3.1 \, (syst)
                  ^{+3.2}_{-2.0} \,(life) \,\%,
\nonumber
\label{eq:BcRatio}
\end{eqnarray}
a rate below LEP sensitivities.
This ratio 
is life\-time dependent, and is shown in Fig.~\ref{fig:life} along with
theoretical predictions \cite{CDFBc}. Two different 
as\-sump\-tions for $\Gamma_{s.l.}(B^+_c \!\rightarrow\! J/\psi \ell^+\nu)$ 
are shown.

\section{Summary and prospects \label{sec:summ}}

We have observed the production of $B^{**}$~states~in~$\bar{p}p$ 
collisions, at a relative rate similar to LEP's.
Pions from $B^{**}$ decays are likely a significant contribution 
to ``same-side'' flavor-tagging methods.
Our~sample~is too limited to unravel the four $B^{**}$ states.
Next year, however, Run II of the Fermilab Tevatron~\cite{CDFup} 
will~begin~where~we~expect 
$20\times$ the luminosity ($\sim\!\!2\,$fb$^{-1}$ in 2 years).
Fully exclusive $B^{**}$ reconstructions should be possible 
with these larger $B$ samples, and the finer mass resolution will aid
in the study of these states.

We have also made the first observation of the $B^+_c$ meson,
and performed a initial survey
of its properties. The increased data of Run II will
enable us to improve all these measurements. This is most notably
the case for the  $B^+_c$ mass, as 
we should be able to fully reconstruct
some of its decay modes.


\end{document}